\documentclass[aps,prl,twocolumn,showpacs,groupedaddress]{revtex4}
\usepackage{graphicx}
\usepackage{color}
\usepackage{epsfig}
\usepackage{dcolumn}   % needed for some tables
\usepackage{bm}        % for math
\usepackage{amssymb}   % for math
\usepackage{ulem}

\newcommand{\bB}{\mathbf{B}}

\newcommand{\bU}{\mathbf{U}}

\begin{document}

\title[]{Transition from large-scale to small-scale dynamo}
\author {Y. Ponty, $^{1}$  F. Plunian,$^{2}$}

\email{Franck.Plunian@ujf-grenoble.fr}

\affiliation{$^1$ Universit\'e de Nice Sophia-Antipolis, CNRS, Observatoire de la C\^ote d'Azur,
BP 4229, Nice cedex 04, France\\$^2$ Institut des Sciences de la Terre, CNRS, Universit\'e Joseph
  Fourier, BP 53, 38041 Grenoble cedex 09, France} 
\begin{abstract}
The dynamo equations are solved numerically with a helical forcing corresponding to the Roberts flow. In the fully turbulent regime the flow behaves as a Roberts flow on long time scales, plus turbulent fluctuations at short time scales. The dynamo onset is controlled by the long time scales of the flow, in agreement
with the former Karlsruhe experimental results.
The dynamo mechanism is governed by a generalized $\alpha$-effect which includes both
usual $\alpha$-effect and turbulent diffusion, plus all higher order effects. 
Beyond the onset we find that this generalized $\alpha$-effect scales as $O(Rm^{-1})$ suggesting the take-over of small-scale dynamo action. This is confirmed by simulations in which dynamo occurs even if the large-scale field is artificially suppressed.
\end{abstract}
\pacs{47.65.-d, 52.65Kj, 91.25Cw}
\maketitle

%\section{Introduction}
The aim of dynamo theory is to understand the physical mechanisms at the origin of magnetic fields in planets and stars. Owing to its complexity it is useful to rely on simple examples for which the dynamo mechanism is well understood. One of them is the one produced by a periodic array of helical vortices. The laminar kinematic dynamo regime has been studied in details by G.O. Roberts \cite{Roberts72}, revealing the two following features.

Firstly the dynamo mechanism
relies on a scale separation between the flow and the magnetic field. 
The largest flow-scale is given by the diameter of one vortex, whereas the magnetic field spreads over an infinite number of them.
This dynamo mechanism is described by two simultaneous effects. The large-scale magnetic field is distorted by the flow resulting into a magnetic field at the scale of one vortex.
This distorted magnetic field and the flow, both at  vortex-scale, combine together to generate a large-scale electromotive force. This large-scale electromotive force induces a large-scale magnetic field, thus closing the loop of the dynamo mechanism. There is even a coefficient of proportionality between the large-scale electromotive force and large-scale magnetic field. It is called $\alpha$
in reference to the ideas developed in the more general context of mean-field theory \cite{Steenbeck66}. 
This dynamo mechanism is said to be large-scale, in reference of the magnetic spectrum which is peaked at the largest scale. 
One decade ago the Roberts dynamo was taken as the starting point for an experimental demonstration of dynamo action \cite{Busse96}.
The experimental results \cite{Stieglitz01} confirmed the theoretical predictions \cite{Tilgner97+}, strongly supporting the large-scale dynamo mechanism.

Secondly, in the Roberts dynamo, the magnetic energy grows at a (slow) diffusive time-scale instead of growing at the (fast) flow turn-over time-scale as expected in turbulent magnetohydrodynamics. Mathematically this results into a magnetic
growthrate $p\rightarrow 0$ in the limit $Rm \rightarrow \infty$, the magnetic Reynolds number being defined as
$Rm= UL / \eta$ where $U$ and $L$ are characteristic flow intensity and length scale, $\eta$ being the magnetic diffusivity. This tendency can be depicted directly from 
\cite{Roberts72} in the curves giving $p$ for different values of $Rm$.
The asymptotic law giving $p$ versus $Rm$ has been derived analytically  \cite{Soward87} and confirmed numerically \cite{Plunian02b}. It was also shown that $\alpha =O(Rm^{-1/2})$, suggesting that the large-scale dynamo mechanism vanishes in the limit of high $Rm$. Recent studies have shown that for other flows, different behaviors of $\alpha$ are also possible \cite{Courvoisier06+}.   

In the context of turbulent dynamos an even steeper scaling
$\alpha =O(Rm^{-1})$ was suggested \cite{Vainshtein92}, due to the nonlinearities occurring in the full dynamo problem composed of the Navier-Stokes and induction equations.
This was confirmed numerically for a flow forcing corresponding to a time-dependent Roberts-like dynamo and for a convective forcing with rotation \cite{Cattaneo96+}.
In that case the dynamo mechanism does not rely on the existence of large magnetic-scales anymore.
The energy transfers, from flow to magnetic field, occur at scales significantly smaller than the largest scale of the system. Small-scale dynamos generally have a higher dynamo onset than the large-scale ones
and are more difficult to obtain at $Pm<1$. In \cite{Ponty0507,Mininni07} advantage was taken from constant flow forcings inducing long-time coherent flows, and then small-scale dynamos have been obtained at $Pm$ down to approximately $10^{-2}$. For non-coherent forcings the numerical evidences are limited to $Pm \ge 1$ so far \cite{Cattaneo96+}, unless other approaches based on hyperviscosity \cite{Schekochihin07} or shell models \cite{Stepanov0608} are used.
Weaker quenching of $\alpha$ have also been found in helical turbulence \cite{Blackman02+},
challenging the previously mentioned results.

%In \cite{Cattaneo96} the authors used three tricks to limit the size of their numerical resolution. First they considered that the flow stays independent of $z$ which is unlikely in a turbulent state.
%Second, instead of varying $Rm$ they varied the intensity of an externally applied magnetic field $B_0$. The goal was to discriminate between two predictions $\alpha \propto B_0^{-2}$ and $\alpha \propto Rm^{-1} B_0^{-2}$. The results agreeing with the second prediction confirmed, indirectly, the $Rm^{-1}$ scaling. Third they considered the case $Pm=1$, where $Pm=\nu / \eta$ is the magnetic Prandtl number, $\nu$ being the fluid viscosity. In planets cores and stars we expect $Pm \ll 1$.

In the present paper we consider the 3D time-dependent problem of Navier-Stokes and induction equations, with a constant forcing corresponding to the Roberts flow geometry. We vary the viscosity in order to explore cases from laminar to fully turbulent flows. For a fully turbulent flow we vary the diffusivity in order to study how the dynamo mechanism varies increasing $Rm$, and eventually determine the transition between large-scale and small-scale dynamo action.

%\section{The model}
%\label{model}
We solve the following set of equations
\begin{eqnarray}
\frac{\partial {\bf U}}{\partial t}&=& - ({\bf U} \cdot \nabla) {\bf U} + (\textbf{B} \cdot \nabla) \textbf{B}
+ \nu \nabla^2 { \bf U}+ {\bf F} \label{Eq_MHDv} \\
\frac{\partial {\bf B}}{\partial t} &=& - ({\bf U} \cdot \nabla) {\bf B} + (\textbf{B} \cdot \nabla) \textbf{U}
+ \eta \nabla^2 {\bf B}
\label{Eq_MHDb}
\end{eqnarray}
where both velocity $\bU$ and magnetic field $\bB$ are assumed to be divergenceless, ${\bf \nabla} \cdot {\bf U} =\nabla \cdot {\bf B} =0$.
The forcing, expressed in a cartesian frame $(\textbf{x},\textbf{y},\textbf{z})$, is given by
\begin{equation}
{\bf  F} = \left(
\sin x \cos y,
- \cos x \sin y, 
 \sqrt{2}~ \sin x \sin y
\right) .
\label{forcing}
\end{equation}
It is force free $\nabla \times {\bf F} = \sqrt{2}{\bf F}$. In the limit of high viscosity $\nu$ and without Lorentz forces, the solution of (\ref{Eq_MHDv}) is given by ${\bf U} = {\bf F} / 2 \nu $,
corresponding to a stationary laminar regime.
Decreasing $\nu$ the non-linear term $({\bf U} \cdot \nabla) {\bf U}$ increases until the flow reaches a turbulent regime. The transition between the laminar and turbulent regime occurs through an oscillatory state as described in \cite{Mininni05}.

For ${\bf U} = {\bf F} / 2 \nu $, the solution of (\ref{Eq_MHDb}) corresponds to the Roberts dynamo solution. 
The large-scale magnetic field $\overline{\bB}$ is then helicoidal and right-handed.
Here $\overline{\bB}$ is defined as the average over the horizontal directions $x$ and $y$.
At a given $z$, it is straight and aligned along one horizontal direction.
The electromotive force $\overline{\cal{E}}=\overline{\textbf{U} \times \textbf{B}}$ shares the same geometry.
In addition the flow symmetries lead to $\overline{\cal{E}}=\alpha \overline{\textbf{B}}$, 
implying the following simple relation
\begin{equation}
	p(k,\eta) = \alpha(k,\eta) k - \eta k^2
	\label{growthrate}
\end{equation}
where the magnetic growthrate $p$ and the ``generalized'' $\alpha$-effect \cite{Soward87}
depend on the magnetic vertical wave number $k$ and the magnetic diffusivity $\eta$.
The ``usual'' $\alpha$-effect and turbulent diffusivity of the mean-field theory \cite{Steenbeck66} 
would correspond to the two first coefficients in the series expansion of $\alpha(k)$ in the limit
$k \rightarrow 0$ \cite{Plunian02b}.

We use a parallelized pseudo-spectral code in a 
periodic box of size $2\pi \times 2\pi \times 4\pi$.
The choice of a box elongated along $z$ corresponds to a minimum magnetic vertical wave number $k^{\text{min}}=0.5$
which we know \cite{Roberts72,Plunian02b} to be more dynamo unstable than $k^{\text{min}}=1$.
Time stepping is done with an exponential forward Euler-Adams-Bashford scheme.

%\section{Dynamo onset}
%\label{Dynamo onset}
%\subsection{Marginal curve}
%\label{marginal_section}

\begin{figure}[htb!]
\begin{center}
\includegraphics[width=0.5\textwidth]{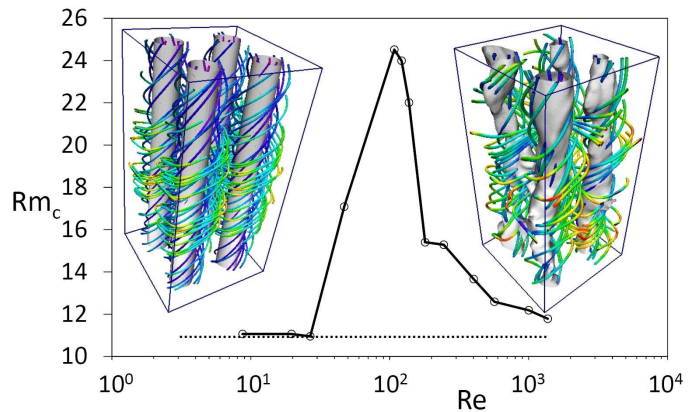}
  \end{center}
\caption{Marginal curve plotted in the $(Re,Rm)$ plane. The insets show snapshots of the flow current-lines and mean (time-averaged) isovalues of the  vorticity $z$-component for two typical regimes, laminar (left) and fully turbulent (right) \cite{vapor}.}
\label{marginal}
\end{figure}

The marginal curve above which dynamo action occurs is plotted in figure \ref{marginal}, with $Re=U_{rms}L_{int}/\nu$ and $Rm=U_{rms}L_{int}/\eta$ \cite{foot1}. 
The numerical values in the simulations are given in table \ref{table_kin}.

\begin{table}
\begin{center}
\begin{tabular}{@{\hspace{0.2cm}}c@{\hspace{0.2cm}}c@{\hspace{0.2cm}}c@{\hspace{0.2cm}}c@{\hspace{0.2cm}}c@{\hspace{0.2cm}}c@{}}
$N_x \times N_y \times N_z$	       &	$\nu$	&	$L_{int}/2\pi$		&	$U_{rms}$			& $\left\langle U\right\rangle$ &	$\eta_c$\\*[0cm]
$64^2$ $\times$ 128	     					 & 1			&	1						      &		0.5					&		0.5                         &	0.28			\\
$64^2$ $\times$ 128	               & 0.6		&	1						      &		0.83				&		0.83                        &	0.47			\\
$64^2$ $\times$ 128	               & 0.4		&	1						      &		1.25				&		1.25                        &	0.71			\\
$64^2$ $\times$ 128                 & 0.3		&	0.88			       	&		1.45				&		1.45                        &	0.73			\\
$64^2$ $\times$ 128                & 0.2		&	0.87		      		&		1.71				&		1.5                         &	0.55			\\
$64^2$ $\times$ 128                & 0.1		&	0.84		      		&		2.03				&		1.7                         &	0.44			\\
$64^2$ $\times$ 128                & 0.09	  &	0.83		      		&		2.09				&		1.7                         &	0.46			\\
$64^2$ $\times$ 128                & 0.08	  &	0.83		       		&		2.12				&		1.67                        &	0.5				\\
$64^2$ $\times$ 128                & 0.06	  &	0.77		       		&		2.22				&		1.53                        &	0.7				\\
$128^2$ $\times$ 256               & 0.05	  &	0.74		        	&		2.62				&		1.55                        &	0.8				\\
$128^2$ $\times$ 256               & 0.03	  &	0.69			        &		2.77				&		1.6                         &	0.882			\\
$128^2$ $\times$ 256               & 0.02	  &	0.65			      	&		2.77				&		1.6                         &	0.9				\\
$256^2$ $\times$ 512					     & 0.01	  &	0.59				      &		2.69				&		1.71    
 &	0.82		 \\
$256^2$ $\times$ 512					     & 0.007	&	0.58				      &		2.63				&		1.72    
 & 0.815			
	\end{tabular}
	\caption{The two first columns correspond to simulation inputs, number of Fourier modes and viscosity. The other columns give the outputs: flow integral scale, $r.m.s.$ velocity, mean-velocity and critical magnetic diffusivity.}
	\label{table_kin}
	\end{center}
\end{table}

At low $Re$ the flow is laminar and stationary, corresponding to the Roberts flow. At high Reynolds numbers the flow is turbulent, though it has a mean (time-averaged) geometry converging towards a Roberts flow. This is illustrated in Figure \ref{marginal} in the two insets. The fact that $Rm_c$ is almost the same for both regimes ($Rm_c \approx 11$, dotted line), suggests that it is the mean-flow which plays the most important role in the field generation, even though it is about $40\%$ less intense than the fluctuations.

This is a drastic difference with other cases like the one obtained with a Von Karman flow forcing \cite{Ponty04, Ponty0507} for which the turbulent onset is always higher than the laminar one. This stresses the robustness of scale-separation dynamos as previously noted \cite{Frick06+}.
In \cite{Mininni05} a higher turbulent onset was found though a Roberts forcing was also used. This discrepancy comes from the fact that in \cite{Mininni05} the periodic box was cubic, corresponding to $k^{\text{min}}=1$. In that case, the onset in the laminar regime is higher by a factor about 4 \cite{Plunian02b}. Presumably at high Reynolds numbers the mean flow is then not strong enough to sustain dynamo action at onset, corresponding to a small-scale dynamo rather than a large-scale one.

At intermediate values of Reynolds number ($Re \approx 10^2$) the dynamo onset is the highest ($Rm_c\approx 25$). The clue to understand this sharp increase of $Rm_c$ lies in the statistical properties of the flow. 
Indeed for such value of $Re$, the mean-flow geometry does not converge
\cite{foot2}. This transition state is characterized by large-scale flow fluctuations, or a lack of long-time coherence, which are known to decrease the dynamo efficiency and then to increase the dynamo onset \cite{Normand03+}.

%\subsection{A large-scale dynamo mechanism}
%\label{mean-field}
At dynamo onset the magnetic field geometry is again helicoidal, right-handed and of $k=0.5$ wave-number, as in the laminar kinematic Roberts dynamo. 
This is a serious hint for a large-scale dynamo mechanism governed by the mean flow.   
Thus
we look for an $\alpha$-tensor defined by
\begin{equation}
	\overline{\cal{E}} = \alpha \overline{\textbf{B}}
	\label{meanemf}
\end{equation}
where $\overline{\cal{E}}$ and $\overline{\textbf{B}}$ are two outputs of the simulation.
As in the Roberts dynamo we find that $|\overline{B}_z|\ll |\overline{B}_x|, |\overline{B}_y|$ and $|\overline{\cal{E}}_z| \ll |\overline{\cal{E}}_x|, |\overline{\cal{E}}_y|$, the $\alpha$-tensor being then reduced to four coefficients.
We find $\left\langle \alpha_{11}\right\rangle\approx \left\langle \alpha_{22}\right\rangle$ and $\left\langle \alpha_{ij}\right\rangle_{i\ne j} \ll \left\langle \alpha_{ii}\right\rangle$. 
Just above or below the onset we find that (\ref{growthrate}) holds for $p=\left\langle p\right\rangle$ and $\alpha=\left\langle \alpha_{11}\right\rangle$, implying $\eta_c=\left\langle \alpha_{11}\right\rangle / k$.
It is an other way to emphasis that the mean-field approach derived by Roberts applies at the onset even in a fully turbulent regime.

%\section{Saturation above dynamo onset}
%\label{above onset}

From now we fix the viscosity $\nu=0.02$ ($Re\approx 570$) and decrease $\eta$ from 0.85 to 0.01
($Rm \in \left[13,1100\right]$).
The number of Fourier modes for all calculations is $128^2 \times 256$
while the length time of resolution is always larger than one diffusion time $\left(2\pi\right)^2/\eta$.

\begin{figure}[htb!]
\begin{center}
\begin{tabular}{@{\hspace{0cm}}c@{\hspace{0cm}}c@{\hspace{0cm}}l@{}}
    \includegraphics[width=0.45\textwidth]{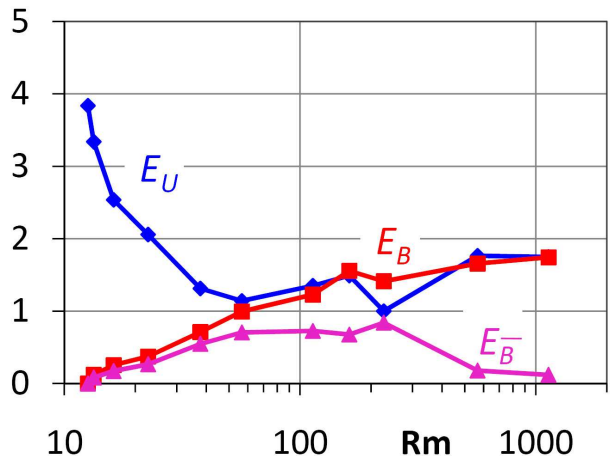}
    \\*[0cm]
  \end{tabular}
  \end{center}
\caption{Kinetic (blue), total magnetic (red) and large-scale magnetic (magenta) energies versus $Rm$ for $\nu=0.02$ ($Re\approx 570$). }
\label{energies_Rm}
\end{figure}

In figure \ref{energies_Rm} the mean kinetic, total and large-scale magnetic energies during the saturation phase are plotted versus $Rm$. 
Here $Rm=U_{rms} L_{int} / \eta$ with values for $U_{rms}$ and $L_{int}$ 
taken from the non magnetic case ($U_{rms}=2.77$ and $L_{int}/2\pi=0.65$).
Increasing $Rm$ we clearly see the tendency towards equipartition between kinetic and magnetic energies, and an increase followed by a decrease of the large-scale magnetic energy. 

\begin{figure}[htb!]
	  \begin{center}
    \includegraphics[width=0.45\textwidth]{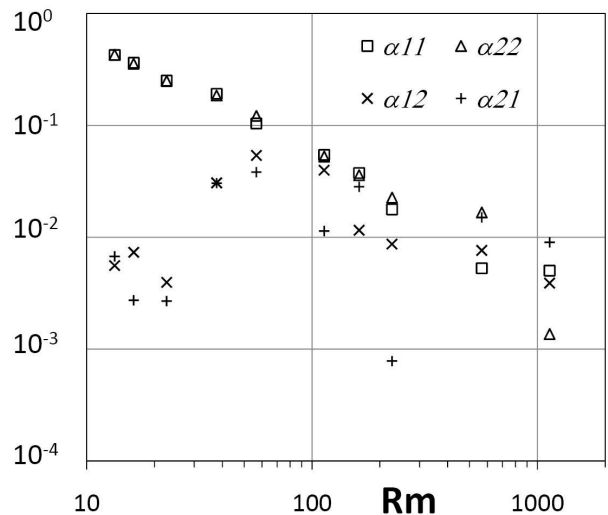}
    \end{center}
\caption{The mean coefficients of the $2 \times 2$ $\alpha$-tensor versus $Rm$, in the saturated state,
for $\nu=0.02$ ($Re\approx 570$). }
\label{alpha_Rm}
\end{figure}

During saturation the $\overline{B}$ and $\cal{E}$ geometries are again helicoidal, right-handed and of $k=0.5$ wave-number. Increasing $Rm$, the correlation between $\overline{B}$ and $\cal{E}$ is weaker than at dynamo onset, implying a somewhat less relevant mean-field interpretation of the results. However, solving (\ref{meanemf}) it is still possible to calculate the $\alpha_{ij}$ coefficients of the $\alpha$-tensor. Their mean values in the saturated state are plotted versus $Rm$ in figure \ref{alpha_Rm} \cite{foot3}.

The diagonal coefficients $\left\langle \alpha_{11}\right\rangle$ and $\left\langle \alpha_{22}\right\rangle$ are found to scale as $O(Rm^{-1})$ over two decades suggesting that the large-scale dynamo mechanism operating at dynamo onset is not the relevant one operating at high $Rm$. The anti-diagonal coefficients $\alpha_{12}$ and $\alpha_{21}$ do not vanish contrary to the kinematic Roberts dynamo, and presumably because of a slight
$z$-dependency of the mean flow.
They first increase versus $Rm$ by a factor 10, and then follow the $O(Rm^{-1})$ scaling for higher $Rm$.
This is reminiscent to the catastrophic quenching in MHD turbulence \cite{Vainshtein92}
though here, the Lorentz forces for the non-linear saturation of the $\alpha$-coefficients mainly occur at the scale of the periodic box, and not at smaller turbulent scales.
We note that this scaling is different from the one found in the kinematic case 
$\alpha=O(Rm^{-1/2})$ \cite{Soward87}. However they are both compatible with (\ref{growthrate}). Indeed in our simulations
$k$ is fixed, whereas in the kinematic case $k=O(Rm^{1/2})$ \cite{Plunian02b}. 

For $Rm < 200$ we find that the nonlinear saturation obeys to a scenario similar to the one described in \cite{Tilgner01} in the laminar regime. The Lorentz force in addition to decreasing the mean-flow intensity modifies its geometry such that the magnetic energy saturates. For $Rm > 50$ this modified mean-flow is able to generate the growth of an additional passive vector field with a phase shifted by $\pi/2$ \cite{Tilgner08+}. For $Rm > 200$ this weakly non-linear scenario does not apply anymore due to too strong nonlinearities. 
However we find that a passive vector field is still growing, suggesting a small-scale dynamo mechanism \cite{Cattaneo09}.

In order to account for such a small-scale dynamo mechanism we solve again equations (\ref{Eq_MHDv}) and (\ref{Eq_MHDb}), but enforcing $\overline{\textbf{B}}=0$ at each time-step \cite{foot4}, in order to suppress any possibility of a large-scale dynamo mechanism. We find a second onset at $Rm \approx 200$,
corresponding to $Pm \approx 0.35$. This shows that provided $Rm$ is high enough the magnetic field grows at small scales, the participation of the large-scale field being sufficiently weak to be neglected in the dynamo process.
Still a $O(Rm^{-1})$ $\alpha$-effect may be calculated provided that small-scale velocity and magnetic field are sufficiently well correlated. A weak large-scale field, enslaved to the small-scale field, may then be generated.

%\section{Discussion}
To conclude, scale separation is confirmed to be a good candidate for liquid metal experiment dynamos
at low $Rm$,
the turbulence having a weak effect on the mean-flow dynamo onset.
In addition we showed that increasing $Rm$, but keeping $Pm<1$, yields to small-scale dynamo action.
Building an apparatus like the Karlsruhe experiment \cite{Stieglitz01} but less constrained would be the cost to explore, above onset, the competition between large-scale and small-scale dynamo modes.\\

We acknowledge fruitful discussions with D. Hughes, A. Gilbert, A. Courvoisier and A. Brandenburg.
YP thanks A. Miniussi for computing design assistance. Computer time was provided by
GENCI in the IDRIS, CINES and CCRT facilities and the Mesocentre SIGAMM machine, hosted by the Observatoire de la C\^ote
d'Azur.


\begin{thebibliography}{}
\bibitem{foot1}
We define $U_{rms}=\left\langle \left(\int 2 E(k) dk\right)^{1/2}\right\rangle$, and
the integral scale $L_{int}/2\pi=\left\langle \int E(k)k^{-1}dk / \int E(k)dk\right\rangle$, where $E(k)$ is the kinetic energy spectrum, and the brackets $\left\langle \cdot \right\rangle$ denote time average.
\bibitem{foot2}
Contrary to integral quantities $U_{rms}, \left\langle U \right\rangle$ and $L_{int}$ which do converge.
\bibitem{foot3}During the kinematic regime the values of the $\alpha_{ij}$ are becoming strongly fluctuating when increasing $Rm$, their mean values being meaningless to calculate.
\bibitem{foot4}In spectral space, it corresponds to canceling the magnetic mode $(k_x=0,k_y=0,k_z=1/2)$.
\end{thebibliography}
\end{document}